\documentclass[aps,prl,nofootinbib,twocolumn,superscriptaddress]{revtex4-1}
\usepackage{euscript}
\usepackage{amssymb}
\usepackage{amsfonts}
\usepackage{amsbsy}
\usepackage{epsfig}
\usepackage{subfigure}
\usepackage{amsthm}
\usepackage{amscd}
\usepackage{amstext}
\usepackage{verbatim}
\usepackage{amsmath}
\usepackage{url}
\usepackage{bm}
\usepackage[pdftex]{hyperref}
\usepackage{color}

\begin{document}
\title{Holographic dissipation prefers the Landau over the Keldysh form}
\author{Yu-Kun Yan}
\email{yanyukun20@mails.ucas.ac.cn}
\affiliation{School of Physical Sciences, University of Chinese Academy of Sciences, Beijing 100049, China}
\author{Shanquan Lan}
\email{lansq@lingnan.edu.cn}
\thanks{Corresponding author}
\affiliation{Department of Physics, Peking University,  Beijing 100871, China}
\affiliation{Department of Physics, Lingnan Normal University, Zhanjiang 524048, China}
\author{Yu Tian}
\email{ytian@ucas.ac.cn}
\thanks{Corresponding author}
\affiliation{School of Physical Sciences, University of Chinese Academy of Sciences, Beijing 100049, China}
\affiliation{Institute of Theoretical Physics, Chinese Academy of Sciences, Beijing 100190, China}
\author{Peng Yang}
\email{yangpeng18@mails.ucas.ac.cn}
\thanks{Corresponding author}
\affiliation{School of Physical Sciences, University of Chinese Academy of Sciences, Beijing 100049, China}
\author{Shunhui Yao}
\email{yaoshunhui15@mails.ucas.ac.cn}
\thanks{Corresponding author}
\affiliation{School of Physical Sciences, University of Chinese Academy of Sciences, Beijing 100049, China}
\author{Hongbao Zhang}
\email{hongbaozhang@bnu.edu.cn}
\thanks{Corresponding author}
\affiliation{Department of Physics, Beijing Normal University, Beijing 100875, China}

\begin{abstract}
Although holographic duality has been regarded as a complementary tool in helping understand the non-equilibrium dynamics of strongly coupled many-body systems, it still remains a remarkable challenge how to confront its predictions quantitatively with the real experimental scenarios. By matching the holographic vortex dynamics with the phenomenological dissipative Gross-Pitaeviskii models, we find that the holographic dissipation mechanism can be well captured by the Landau form rather than the Keldysh one, although the latter is much more widely used in numerical simulations. Our finding is expected to open up novel avenues for facilitating the quantitative test of the holographic predictions against the upcoming experimental data. Our result also provides a prime example how holographic duality can help select proper phenomenological models to describe far-from-equilibrium nonlinear dynamics beyond the hydrodynamic regime.
\end{abstract}

\maketitle

\emph{Introduction}.---For the non-equilibrium dynamics of strongly interacting quantum systems where the quasi-particle picture does not apply and the perturbation method fails, developing its theoretical description remains an important task\cite{Sachdev2011,LiuH2020}. Gratefully, holographic duality\cite{Maldacena97,Gubser98,Witten98}, also known as anti-de-Sitter space/conformal field theory
correspondence, has provided a powerful insight into the universal behaviors of strongly coupled dynamics through the classical theory of gravity with one additional dimension. In particular, a variety of bottom-up gravitational models have been proposed to address the strongly correlated condensed matter systems\cite{Hartnoll2008,Hartnoll2018,Herzog2009,McGreevy2010,Adams2013,Solana2014,Zaanen2015}. But nevertheless, associated with these bottom-up holographic models, there exists a significant deficiency, namely, the effective dual boundary descriptions are generically unknown, which makes it a notoriously difficult challenge to compare the holographic prediction with the experimental data.

Among others, the dynamics of the quantized vortices in superfluids, which plays a vital role in the fascinating non-equilibrium quantum turbulence, have recently become amenable to being engineered at finite temperature in a controllable manner due to the great experimental advances in cold atom gases\cite{Sachkou2019,kwon2021}. In contrast to classical turbulence in normal fluids, which can be well described by dissipative hydrodynamics, quantum turbulence in superfluids exits the hydrodynamic regime due to the very presence of the quantized vortices. It is thus urgent to construct an effective boundary description of holographic superfluids, which provides a complete description--valid at all scales--of the superfluid dynamics, including the vortex dynamics.

On the other hand, different from the holographic duality, which provides a universal first principles description of the irreversible finite temperature dissipation in terms of the excitations absorbed by the bulk black holes, the conventional approach to incorporate the dissipation in superfluids is essentially phenomenological. The dissipation terms in the different phenomenological models will give rise to different predictions as it should be the case. Therefore it is desirable to resort to a first principles calculation to help select which phenomenological model is a proper one.

This {\it Letter} intends to serve as such one stone which attempts to kill the above two birds by matching the two available phenomenological dissipative Gross-Pitaevskii equations with the holographic vortex dynamics. As a
result, we find that the dissipation mechanism in our holographic superfluid can be well described by the dissipative Gross-Pitaevskii equation with the dissipation given by the Landau form rather than the Keldysh one, although the latter is much more commonly used in numerical simulations of superfluid dynamics. Compared to the previous progresses such as made in \cite{Ammon2020, Arean2021}, which nonetheless restricts mainly within the equilibrium state or the near-equilibrium hydrodynamic regime, our work presents us a prime example how holographic duality can also help select proper phenomenological models to describe the far-from-equilibrium nonlinear dynamics beyond the hydrodynamic regime. On the other hand,
with our finding, the holographic superfluid model with four bulk dynamical variables and one adjustable boundary value can be described effectively by only one dynamical variable with three adjusted parameters in one less dimension, which will make the quantitative comparison of the holographic predictions with the real upcoming experimental data much easier and much more efficient.

\emph{Holographic superfluids model and dissipative Gross-Pitaevskii models}.---In the probe limit, where the holographic superfluid is implemented by the Abelian Higgs model with the Lagrangian density given by\cite{Hartnoll2008,Herzog2009}
\begin{eqnarray}
\mathcal{L}&=&-\frac{1}{4}F^{\mu\nu}F_{\mu\nu}-|D_\mu\Phi|^2-m^2|\Phi|^2,
\end{eqnarray}
on top of the (3+1) dimensional planar Schwarzschild AdS black hole in the Eddington-Finkelstein coordinates
\begin{equation}
ds^2=\frac{L_{AdS}^2}{z^2}(-f(z)dt^2+dx^2+dy^2-2dtdz),
\end{equation}
where $D_\mu=\nabla_\mu-iA_\mu$ and $f(z)=1-(z/z_h)^3$ with $z_h$ the location of the black hole horizon. The corresponding dynamics is governed by the following equations of motion
\begin{equation}
\begin{split}
&\nabla_{\mu}F^{\mu\nu}=i(\Phi^*D^{\nu}\Phi-\Phi(D^{\nu}\Phi)^*),\\
&D_{\mu}D^\mu\Phi-m^2\Phi=0,
\end{split}\label{eoms}
\end{equation}
where the asterisk denotes the complex conjugation.

By holography, the temperature of the dual boundary system is given by the Hawking temperature $\tilde{T}=3/(4\pi z_h)$, and the chemical potential is related to the boundary data of the bulk field $A_t$ as  $\tilde{\mu}=A_t|_{z=0}$. Due to the scaling symmetry, one can set $z_h=1$ once and for all. Accordingly, it turns out when the chemical potential is higher than the critical value $\tilde{\mu}_c=4.064$, the bulk complex scalar field will spontaneously condense, which signals the transition to the superfluid phase on the boundary. The corresponding order parameter $\psi$ can be read off from the boundary data of $\Phi$ according to the holographic dictionary. It is noteworthy that holography provides a natural built-in mechanism to account for the irreversible finite temperature dissipation by geometrizing the excitations absorbed by the black hole.

Different from the above holographic model of superfluids, the conventional phenomenological models have significant limiations and shortcomings, where the dissipation is essentially put in by hand. As to the Bose-Einstein condensates (BEC) in the dilute cold atom gases at nearly zero temperature, the behaviour of order parameter $\psi$ can be successfully described by Gross-Pitaevskii equation (GPE)
\cite{GP}. However, GPE cannot describe BEC at finite temperature. In order to account for the finite temperature effect, one is required to introduce the dissipative terms.
For our purpose, we consider two such dissipative Gross-Pitaevskii equations (DGPEs), which can be written in the dimensionless form as follows 
\begin{eqnarray}
&\partial_t\psi&=-\frac{(i+\gamma)}{2\tau}(-\nabla^2\psi+2\mu(|\psi|^2-1)\psi),\label{dgp1}\\
&\partial_t\psi&+i\lambda\psi\partial_t|\psi|^2=-\frac{i}{2\tau}\left[(-\nabla^2\psi+2\mu(|\psi|^2-1)\psi)\right].\label{dgp2}
\end{eqnarray}
Here the parameter $\tau$ controls the characteristic time scale of dynamics, and $\mu$ is the chemical potential, from which the dimensionless healing length is given by $\xi=(2\mu)^{-1/2}$. The dissipative parameter $\gamma$ in Eq.(\ref{dgp1}) is suspected to be determined by the Keldysh self-energy through the fluctuation-dissipation theorem\cite{Stoof1997,Stoof1999,Duine2001}. So we call this equation as KGPE. On the other hand, we denote Eq.(\ref{dgp2}) with $\lambda$ the dissipative parameter as LGPE, as it was phenomenologically motivated by Landau's requirement that the second law of thermodynamics hold in his two fluid model for superfluidity\cite{Landau,Carlson1996}.

\begin{figure}[tbp]
\centering
\subfigure[]{
  \includegraphics[width=8.5cm,height=5cm]{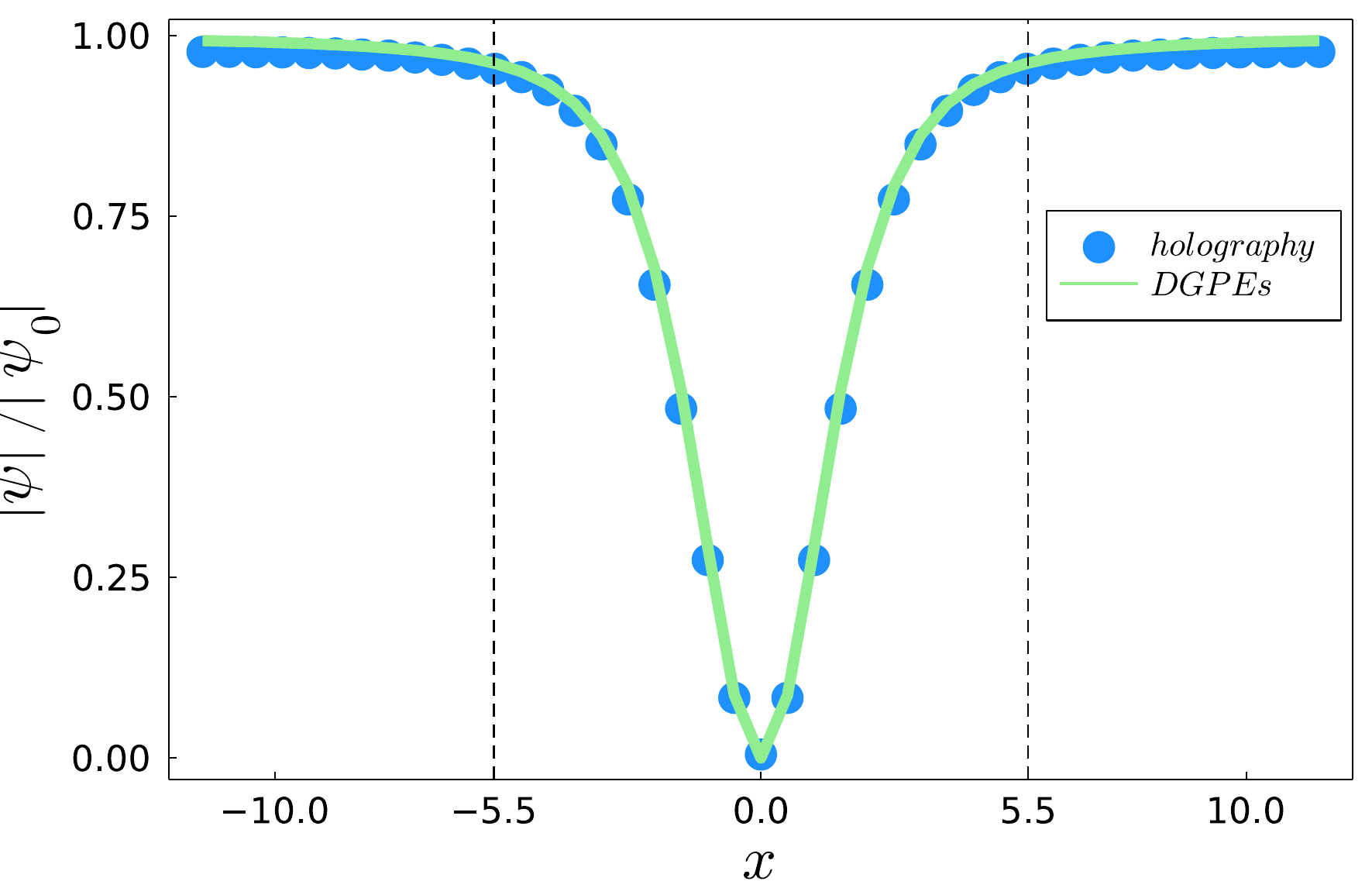}}
\subfigure[]{
  \includegraphics[width=8.5cm,height=5cm]{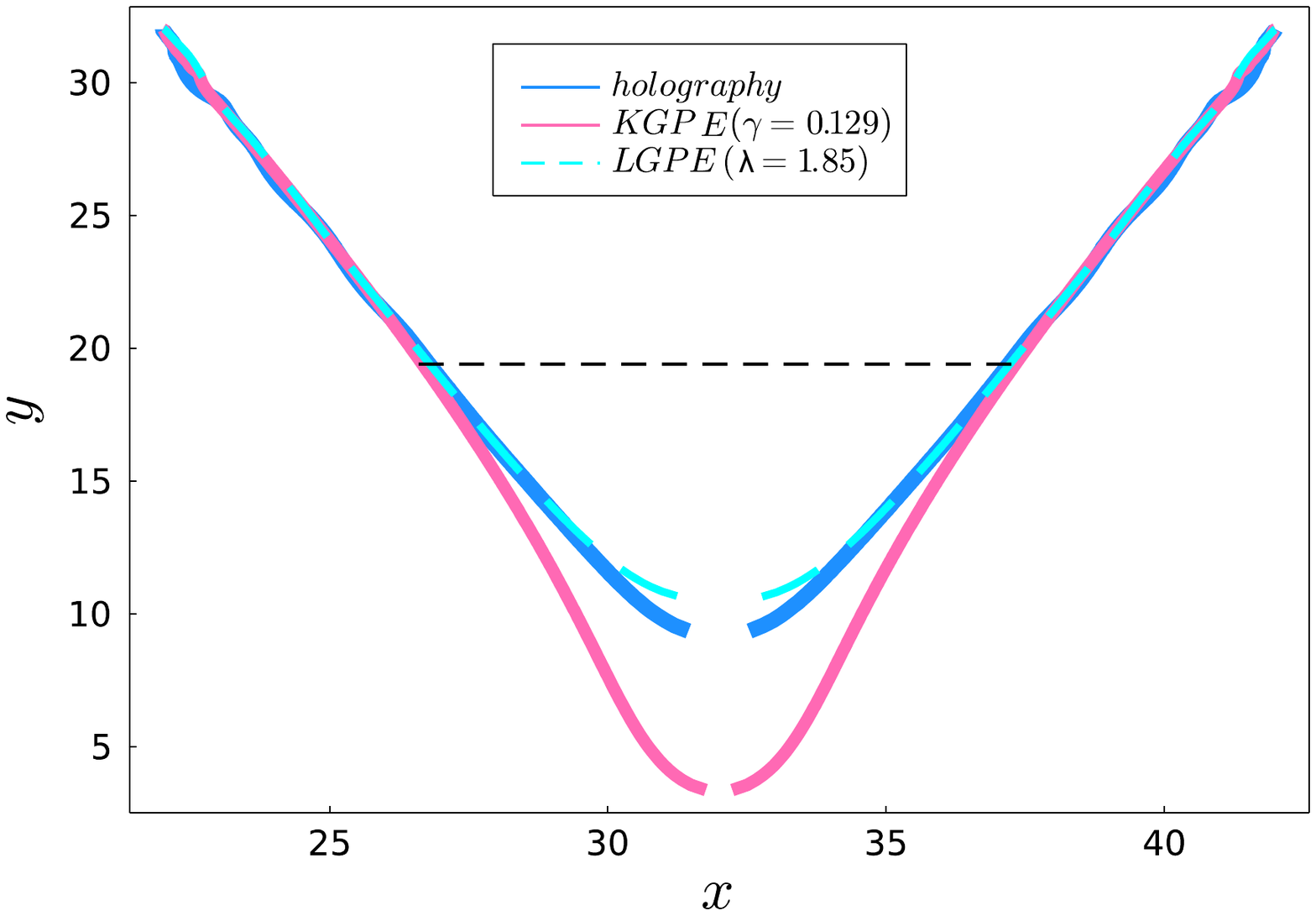}}
\caption{The matching results between the holographic superfluid at $\tilde{\mu}=4.5$ and DGPEs. In (a), the normalized condensate profile of a single static holographic vortex is well fitted by both DGPEs, where the black dotted line is used to identify the vortex size. In (b), the holographic vortex dipole trajectory is fitted by both KGPE and LGPE, where the black dotted line indicates the location where the vortex dipole get contacted with each other.}\label{figsm}
\end{figure}
\begin{figure}[tbp]
\centering
\subfigure[]{
  \includegraphics[width=8cm,height=5cm]{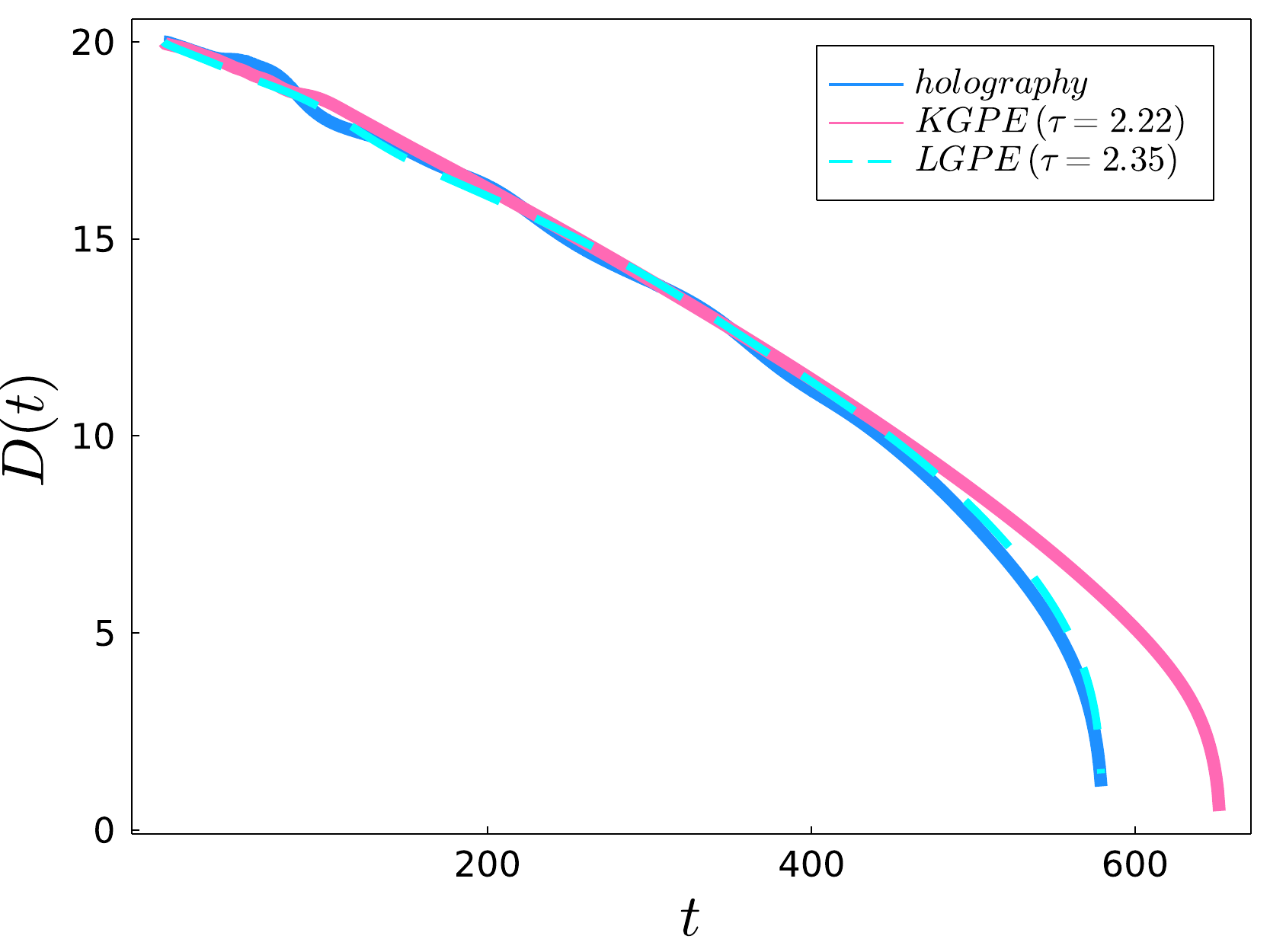}}
\subfigure[]{
  \includegraphics[width=8cm,height=5cm]{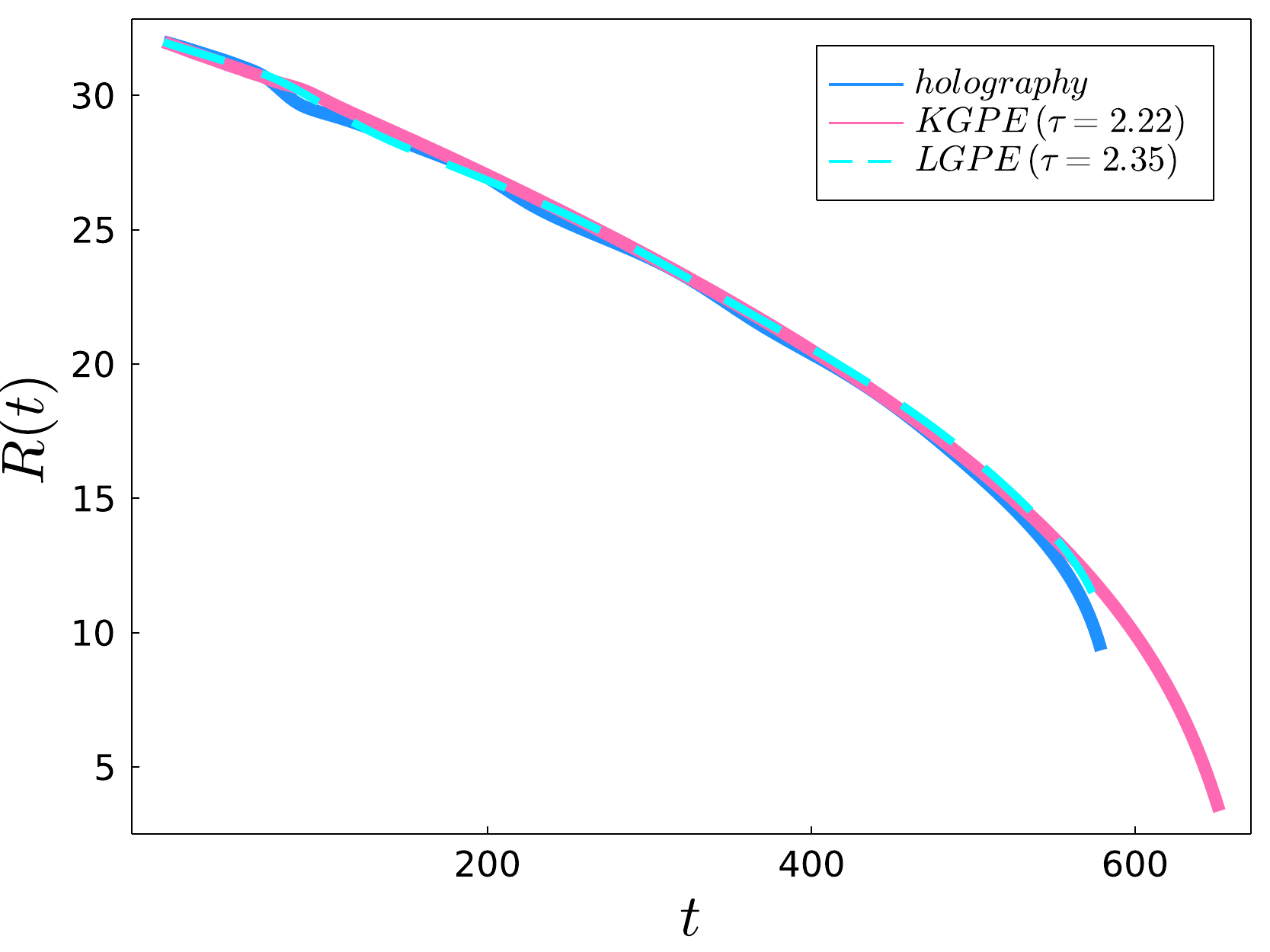}}
\caption{The matching results for the temporal evolution of the relative distance and the center position of the holographic vortex dipole at $\tilde{\mu}=4.5$ by DGPEs in (a) and (b), respectively.}\label{fig4}
\end{figure}

\emph{Matching procedure and relevant results}.---In order to quantify how well the above two models serve as a phenomenological description of the holographic vortex dynamics, we need a matching procedure. In \cite{Ewerz2020} the authors proposed such a procedure, but used an invalid evolution scheme in holography (as explained in Supplemental Material) and then made an unreliable claim that the holographic vortex dynamics was well described by KGPE. Here we use the correct evolution scheme while still employing a similar procedure. Namely we first determine the healing length in both models by fitting the order parameter profile for the holographic vortex of winding number $1$ with the form $|\psi|^2\propto\frac{\textbf{x}^2}{2\xi^2+\textbf{x}^2}$. Then we intend to fit the holographic vortex dipole trajectory by adjusting the corresponding dissipation parameter. Finally, the parameter $\tau$ is fixed by tracking the real time evolution of the vortex dipole.
\begin{table}[tbp]
\caption{The best fitting parameters in DGPEs for the holographic superfluid at $\tilde{\mu}=4.5$ and $\tilde{\mu}=6$.}
\label{tab1}%
\begin{ruledtabular}
\begin{tabular}{cccccc}
\ DGPEs & $\tilde{\mu}$ & $\mu$ & $\lambda$ & $\gamma$ & $\tau$ \\ \hline
LGPE & 4.5 & 0.50 & 1.85 & / & 2.35\\
LGPE & 6  & 2.61 & 1.51 & / & 4.70 \\
KGPE & 4.5 & 0.50 & / & 0.129  & 2.22 \\
KGPE & 6 & 2.61 & / & 0.085 & 4.50\\
\end{tabular}%
\end{ruledtabular}
\end{table}
We demonstrate our relevant results by focusing on a typical example, namely the holographic superfluid at $\tilde{\mu}=4.5$. As illustrated in FIG.\ref{figsm}a, the resulting holographic vortex can be well fitted by both models with the same healing length $\xi=1.0$. On the other hand, as shown in FIG.\ref{figsm}b, the corresponding holographic trajectory can be better modeled by LGPE with $\lambda=1.85$ till the vortex dipole annihilation than KGPE, which starts to display an apparent deviation from the holographic behavior when the vortex dipole get contacted with each other. Similarly, as one can see in FIG.\ref{fig4}, the real time evolution of the holographic vortex dipole can also be better captured by LGPE with $\tau=2.35$ all the way to the annihilation stage than KGPE, which fails to describe the real time dynamics of the vortex dipole when get closed to each other. Similar matching results apply to the holographic superfluid at other chemical potentials. Here we only list the resulting best fitting parameters in TABLE.\ref{tab1} for $\tilde{\mu}=4.5$ and $\tilde{\mu}=6$.
\begin{figure}[tbp]
\centering
\includegraphics[width=8cm]{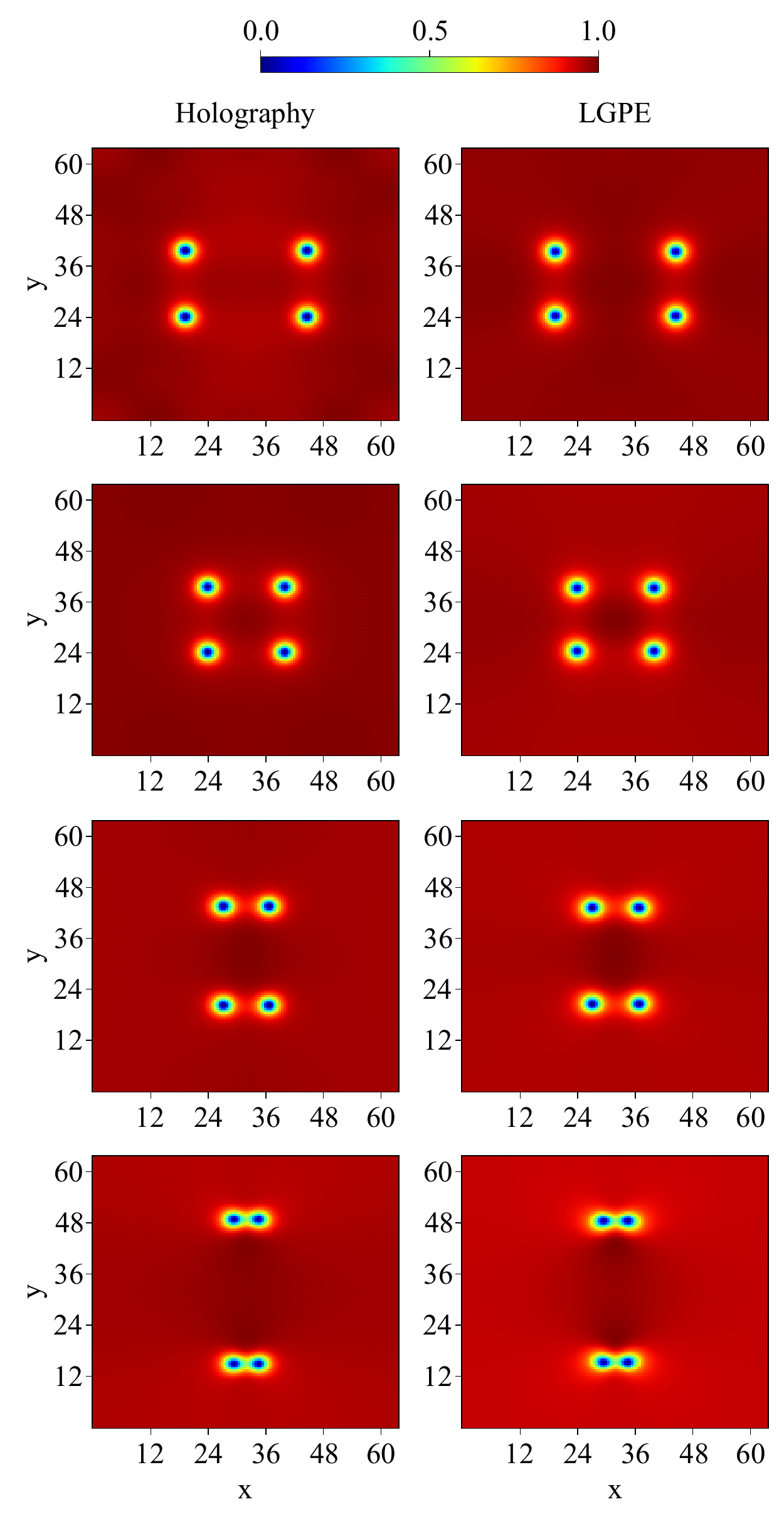}
\caption{The density plot of the normalized condensate
$|\psi|^2/|\psi_0|^2$ for the head on collision of the vortex dipoles in holographic superfluid at $\tilde{\mu}=4.5$ (left) and the matched
LGPE (right), which displays good agreement with each other. The top panel is for the initial stage, where the left and right moving vortex dipoles are prepared. The second panel is for the intermediate collision stage. The third panel denotes the newly formed vortex dipoles moving away from each other. The bottom panel is for the final annihilation stage.}\label{headon_hoL}
\end{figure}
\begin{figure}[tbp]
\centering
\subfigure[]{
  \includegraphics[width=8cm,height=5cm]{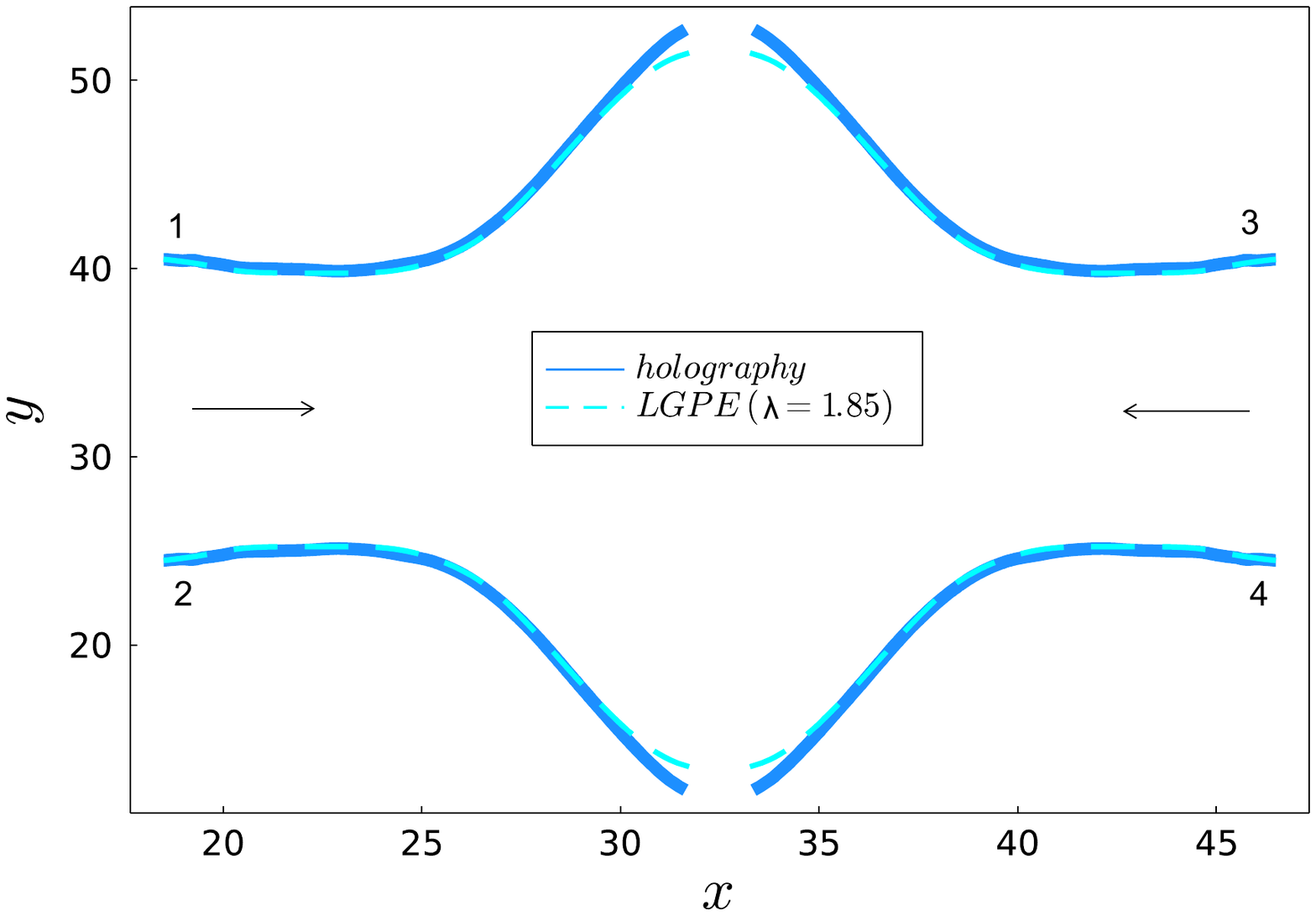}}
\subfigure[]{
  \includegraphics[width=8cm,height=5cm]{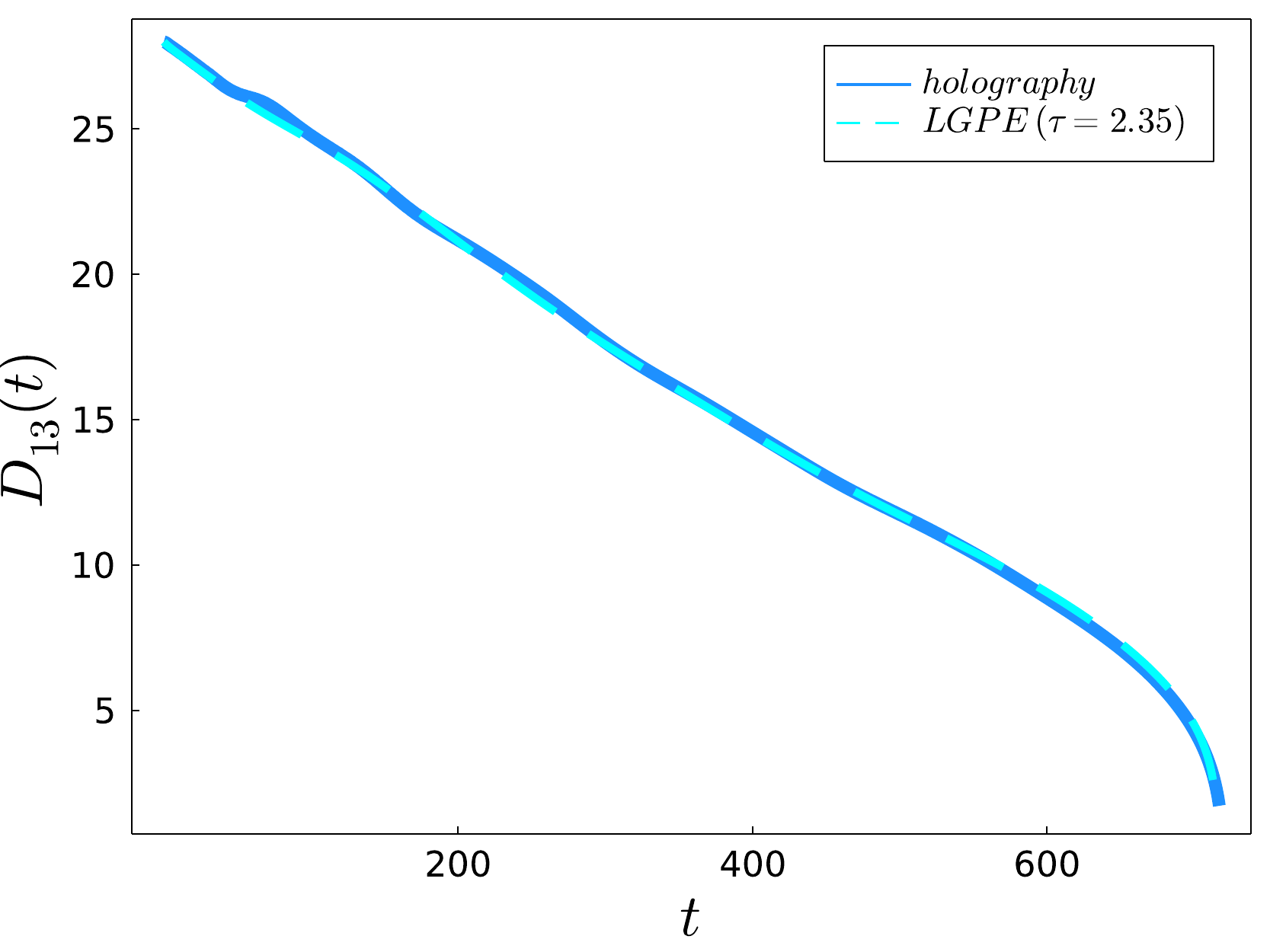}}
\caption{The good agreement between the holographic simulation at $\tilde{\mu}=4.5$ and the matched LGPE on the trajectories of vortices in (a) and the temporal evolution of the relative distance between the vortex `1' from the right moving vortex dipole  and the antivortex `3' from the left moving vortex dipole in (b) during the head on collision of the two vortex dipoles.}\label{headon}
\end{figure}

To substantiate the aforematched LGPE as the effective description of holographic vortex dynamics, we are left to check its generalization capability in other scenarios involving the vortex dynamics. As a demonstration, we examine the head on collision of two vortex dipoles in light of the experimental setup prepared in \cite{kwon2021}. To be more specific, with the matched parameters in TABLE.\ref{tab1}, we compare the numerical result from LGPE and that from our holographic simulation. We first present the four different stages for the head on collision along the horizontal direction by density plot of the condensate in FIG.\ref{headon_hoL}, where the vortex manifest themselves at the locations of zero density. After the collision, the vortex(anti-vortex) from the right moving vortex dipole is seen to recombine with the anti-vortex(vortex) from the left moving one, leading to the formation of new vortex dipoles. Then the new vortex dipoles move away from each other with one marching up and the other marching down. Eventually, both vortex dipoles get annihilated. As illustrated in FIG.\ref{headon_hoL}, both results are in good agreement with each other. We further confirm this in FIG.\ref{headon} by tracking the motion of the involved four vortices. As one can see, The result from our holographic simulation still displays good agreement with that from LGPE till the annihilation of vortex dipoles. Actually, as demonstrated in Supplemental Material, such good agreement between the matched LGPE and the holographic superfluid is also confirmed in more complicated scenarios such as the oblique collision of two vortex dipoles and the random motion of six vortices. This indicates that the matched LGPE can serve as an effective description of holographic vortex dynamics.

\emph{Discussions}.---By fitting the two available phenomenological DGPEs with the holographic superfluid model, we find that the holographic vortex dipole dynamics can be well matched by LGPE all the way down to the vortex dipole annihilation rather than KGPE, which matches up with our holographic data only when the vortex dipole are far apart from each other and displays an apparent deviation when the vortex dipole get close to each other. Although KGPE are much more widely used to attempt modeling the finite temperature BEC for decades than LGPE, actually the linear response theory of KGPE suffers from a serious defect, which, to our best knowledge, has not been noticed before (see Supplemental Material). Together with the observation that LGPE displays a better consistence with holographic vortex dipole dynamics than KGPE, we are convinced that the reasonable phenomenological model for our holographic superfluid should be LGPE rather than KGPE. Our finding also invalidates the claim made recently by the authors in \cite{Ewerz2020} that the holographic vortex dipole dynamics can be well fitted by KGPE. As detailed in Supplemental Material, their wrong result arises from the fact that a defective evolution scheme for the holographic numerical simulation is invoked therein.  In this regard, our result presents a prime example how a proper phenomenological dissipative model can be selected through the lens of holography to describe the far-from-equilibrium nonlinear dynamics beyond the hydrodynamic regime. We further consolidate LGPE as an effective description of holographic vortex dynamics by demonstrating its remarkable generalization capability in more complicated scenarios.

On the other hand, although the holographic superfluid model is superior to those phenomenological models such as DGPEs in the sense that it offers a first principles description of non-equilibrium dissipative dynamics at finite temperature,
not only do  DGPEs live in one less dimension, but also involve only one dynamical variable. Thus it is much easier and much more efficient for one to perform a large scale of  numerical simulations using DGPEs once the undetermined parameters are fixed.  Now according to our matching result, LGPE is selected by holography to serve as the appropriate phenomenological model for the vortex dynamics, so we can use it to greatly facilitate the quantitative confrontation of our holographic predictions with the real experimental data.  In particular, with the recent experimental progress in the vortex dynamics\cite{Sachkou2019,kwon2021}, we expect our results can be verified by the upcoming experiments in the future.

Last but not least, it is important to go beyond the probe limit taken in this Letter. This is tantamount to including the backreaction of the matter fields onto the bulk metric. With this, one can explore the interaction between the stress tensor and the charge current and see how the superfluid component affects the dynamics of normal component. In particular, it is interesting to check whether the full dynamics can also be matched by the effective field theory approach to the superfluid dynamics at finite temperature \cite{Mitra2021}.

\emph{Acknowledgements}.---We would like to thank Carlo Ewerz for his kind clarifications regarding his work, and Huabi Zeng for his helpful discussions. This work is supported in part by NSFC under Grant Nos.11975235, 12005088, 12035016, and 12075026.


\onecolumngrid
\newpage

\section{supplemental material}
\subsection{Numerical details about holographic evolution schemes and error estimates}
 In our holographic simulations, we set $L_{AdS}=1$. In addition, for simplicity but without loss of generality, we take $m^2=-2$. With the choice of the axial gauge $A_z=0$, the above equations of motion can be decomposed into the constraint equation
\begin{equation}
0=-\partial_z^2 A_t+\partial_z\partial\cdot\textbf{A}-2\textrm{Im}(\phi^*\partial_z\phi),\label{constrain}
\end{equation}
and the evolution equations
\begin{eqnarray}
\partial_t \partial_z \phi&=&\partial_z(\frac{f(z)}{2}\partial_z\phi)+\frac{1}{2}\partial^2\phi-i\textbf{A}
\cdot\partial\phi +iA_t\partial_z\phi
-\frac{i}{2}(\partial\cdot\textbf{A}-\partial_zA_t)\phi
-\frac{1}{2}(z+\textbf{A}^2)\phi, \label{dphi}\\
\partial_t \partial_z \textbf{A}&=&\partial_z(\frac{f(z)}{2}\partial_z\textbf{A})-|\phi|^2\textbf{A}+\textrm{Im}(\phi^*\partial\phi)
+\frac{1}{2}\left[\partial\partial_zA_t+\partial^2\textbf{A} -\partial\partial \cdot\textbf{A}\right], \label{eqeom} \\
\partial_t\partial_z A_t&=&\partial^2A_t-\partial_t\partial\cdot\textbf{A}+f(z)\partial_z\partial\cdot\textbf{A} -2A_t|\phi|^2
+2\textrm{Im}(\phi^*\partial_t\phi)-2f(z)\textrm{Im}(\phi^*\partial_z\phi),
\label{eqAt}
\end{eqnarray}
where $\phi=\Phi/z$ and $\textbf{A}=(A_x, A_y)$.
To perform a full non-linear numerical simulation of the above equaitons of motion, we would first like to impose the boundary conditions onto the other bulk fields as follows
\begin{equation}
\phi|_{z=0}=0, \ \ A_t|_{z=0}=\tilde{\mu}, \ \ \textbf{A}|_{z=0}=0.\label{bcs}
\end{equation}
We evolve $\phi$ and $\mathbf{A}$ by Eq.(\ref{dphi}) and Eq.(\ref{eqeom}) but solve $A_t$ by the constraint equation with an extra boundary condition $\partial_z A_t|_{z=0}=-\rho$, where $\rho$ corresponds to the boundary particle number density. The dynamics of $\rho$ is controlled by the current conservation law, which is simply Eq.(\ref{eqAt}) evaluated at the boundary. Thus by implementing the current conservation during our evolution, Eq.(\ref{eqAt}) is guaranteed to hold automatically elsewhere in the whole bulk. Different from ours, the authors in \cite{Ewerz2020} take $A_t|_{z=z_h}=0$ as the extra boundary condition and simply disregard Eq.(\ref{eqAt}). In order to compare our numerical simulation with that in \cite{Ewerz2020}, we prepare the initial data by essentially the same strategy albeit maybe a little bit rough. Namely, the initial $\phi$ is constructed by multiplying the static and homogeneous background $\phi(z)$ with a vortex dipole profile in the \textbf{x} plane. In addition, the initial value for $\textbf{A}$ is set to zero. As a result,  the initial $A_t$ with $A_t|_{z=z_h}=0$ and $A_t|_{z=0}=\tilde{\mu}$ can be obtained by solving the constraint equation followed by the initial $\rho$ extracted through $\rho=-\partial_z A_t|_{z=0}$.

\begin{figure}[tbp]
\centering
\subfigure[]{
\includegraphics[width=4cm]{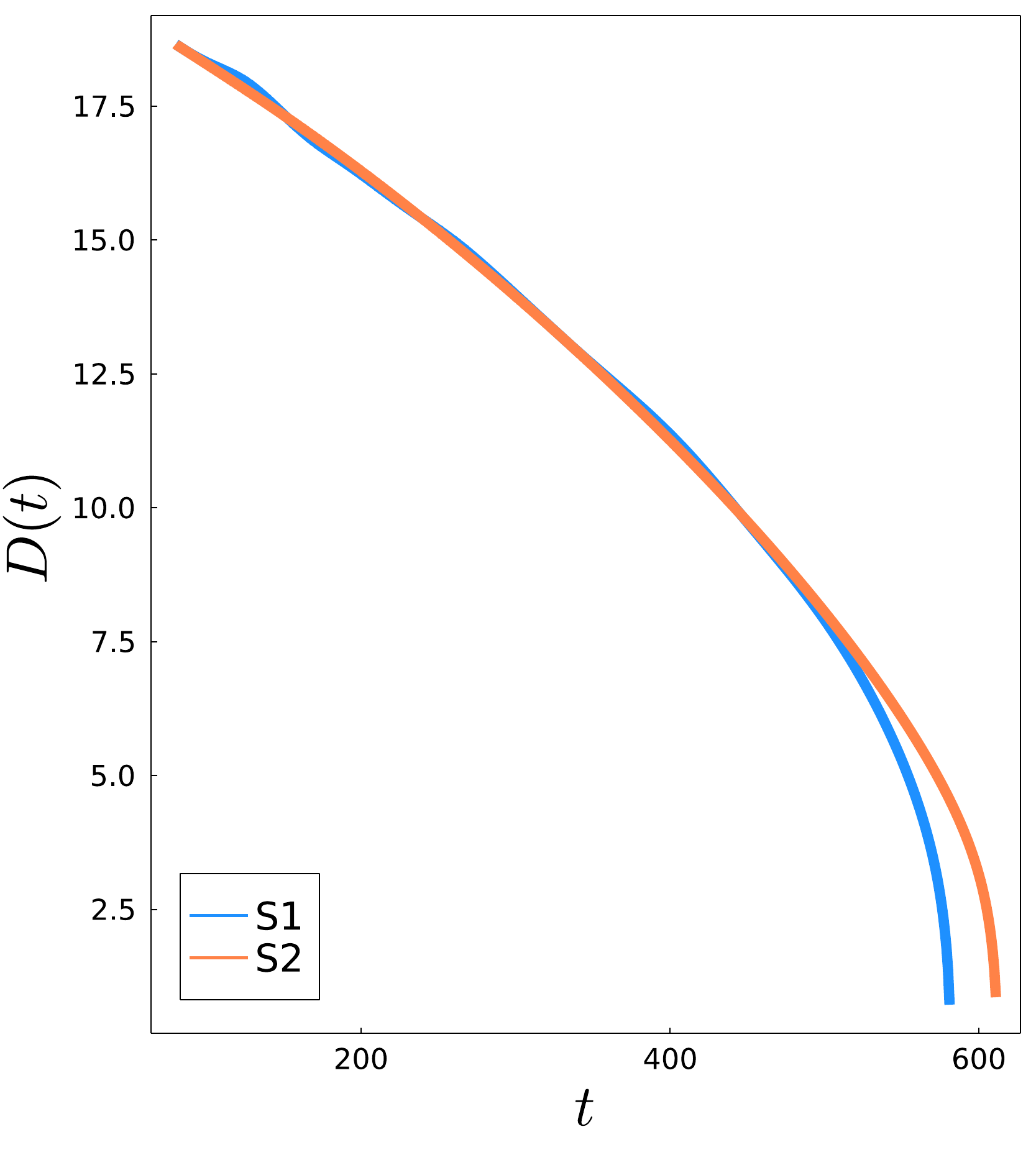}}
\subfigure[]{
\includegraphics[width=4cm]{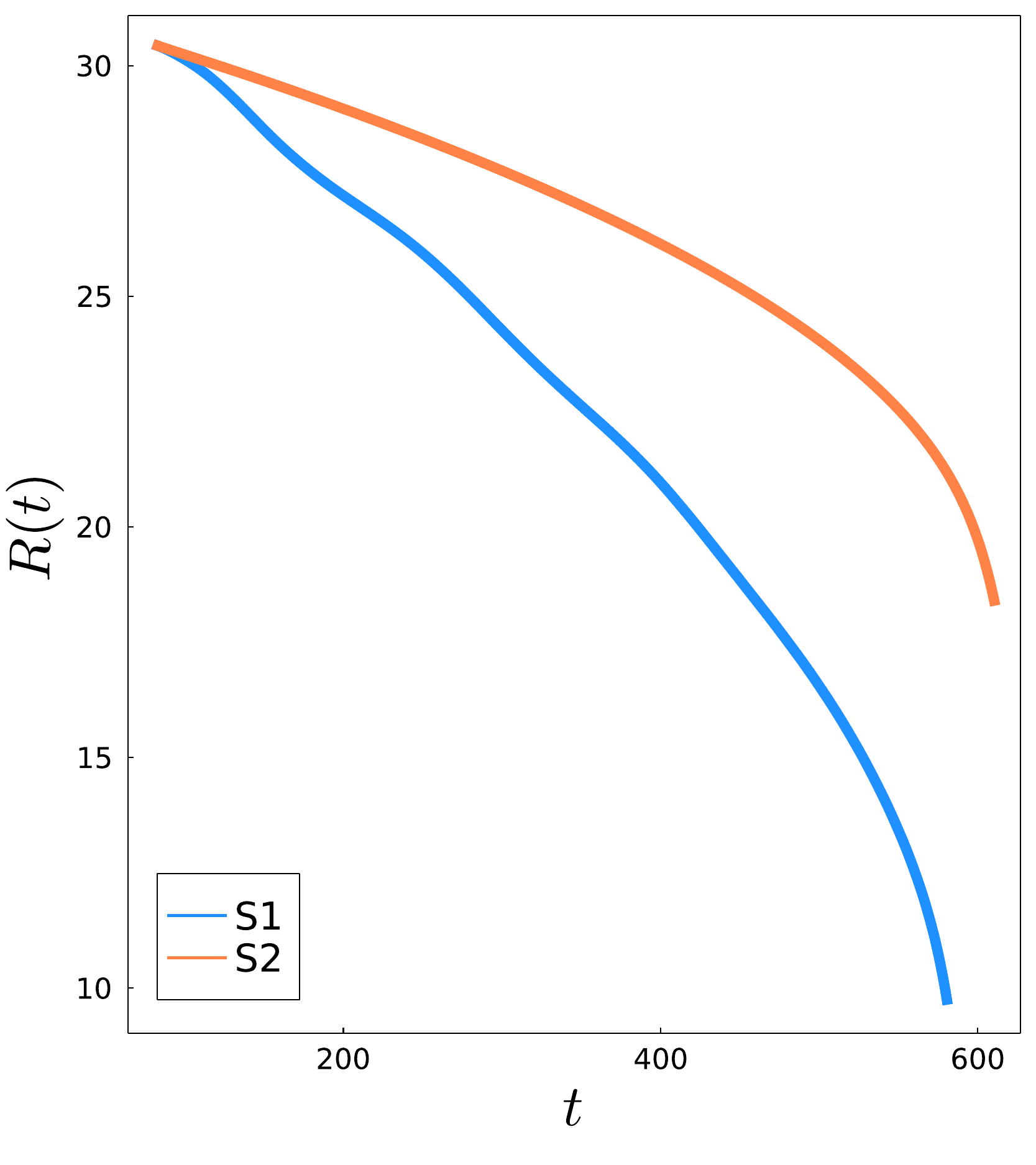}}\\
\subfigure[]{
\includegraphics[width=8.5cm,height=5cm]{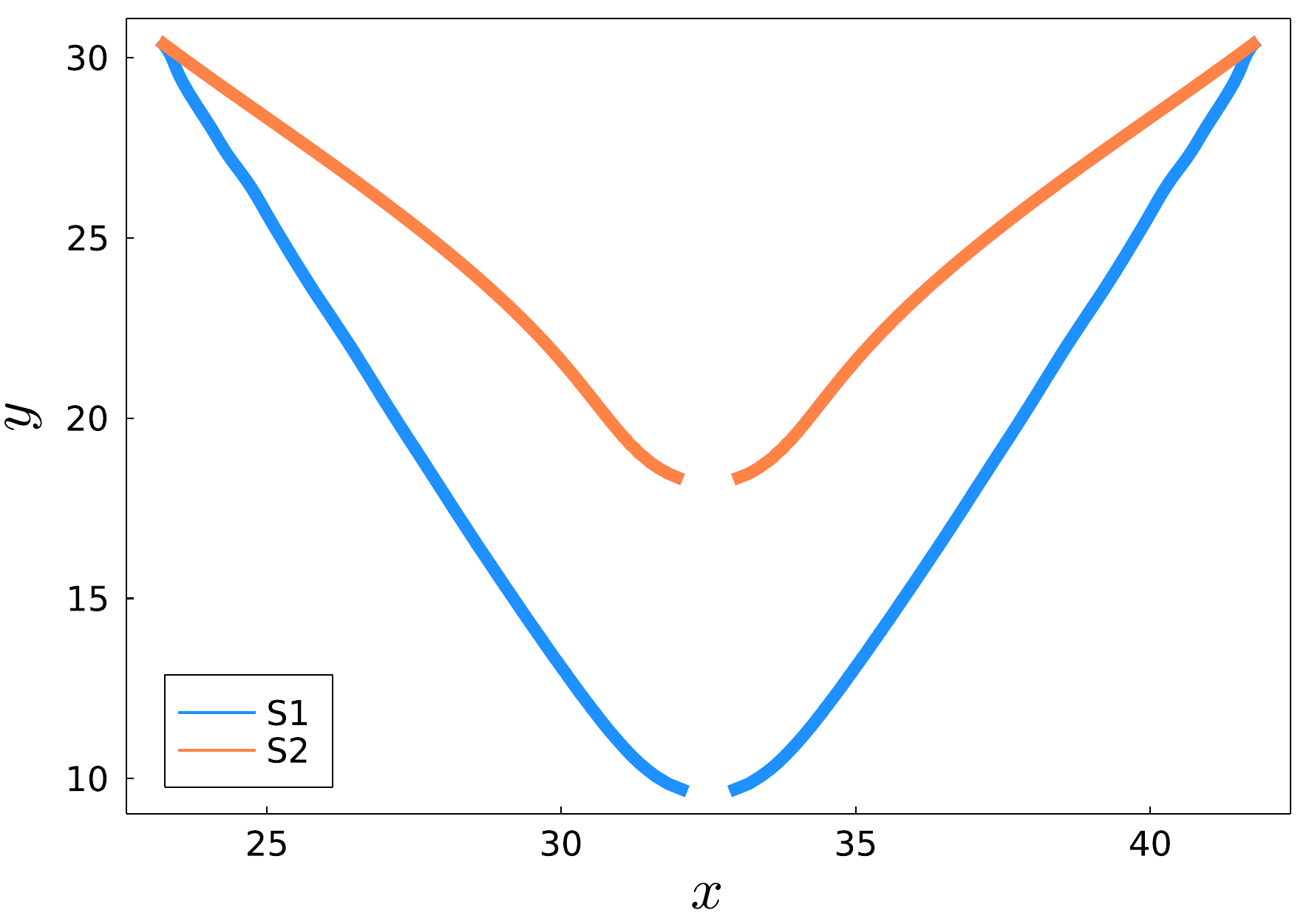}}
\caption{The holographic vortex dipole motion from \textbf{S1} and \textbf{S2} with the same initial data for $\tilde{\mu}=4.5$, where (a) and (b) show the temporal evolution of the relative distance $D$ and the center position $R$ of the vortex dipole, with the resulting trajectories displayed in (c). } \label{fig2}
\end{figure}

We like to demonstrate the numerical results for the vortex dipole motion by our scheme (\textbf{S1}) and (\textbf{S2}) devised in \cite{Ewerz2020} with the exactly same initial data in Fig.\ref{fig2}. As one can see from (a), the relative distance $D$ of the vortex dipole starts to exhibit different behaviors at the vortex dipole annihilation stage such that it takes relatively more time for their vortex dipole to annihilate than ours. Thus in comparison with ours, their scheme produces a kind of repulsive force in addition to the familiar Magnus force between the vortex dipole when closed to each other. Moreover, as illustrated in (b), it is from the very beginning that the center position $R$ of their vortex dipole moves forward in a totally different pace from ours. As a result, the trajectory for their vortex dipole is quite distinct from ours, which is displayed in (c).

The viability of the evolution schemes can be examined by checking the degree of violation of Eq.(\ref{eqAt}) and its convergence during the evolution. As such, we define the error function by
\begin{equation}
\mathcal{E}=\partial_t (S-\partial_zA_t),
\end{equation}
where $S$ is solved numerically according to $\partial_tS=\mathcal{F}_{A_t}$ with $\mathcal{F}_{A_t}$ representing the right side of Eq.(\ref{eqAt}). 
\begin{figure}[h]
\centering
\includegraphics[width=8cm]{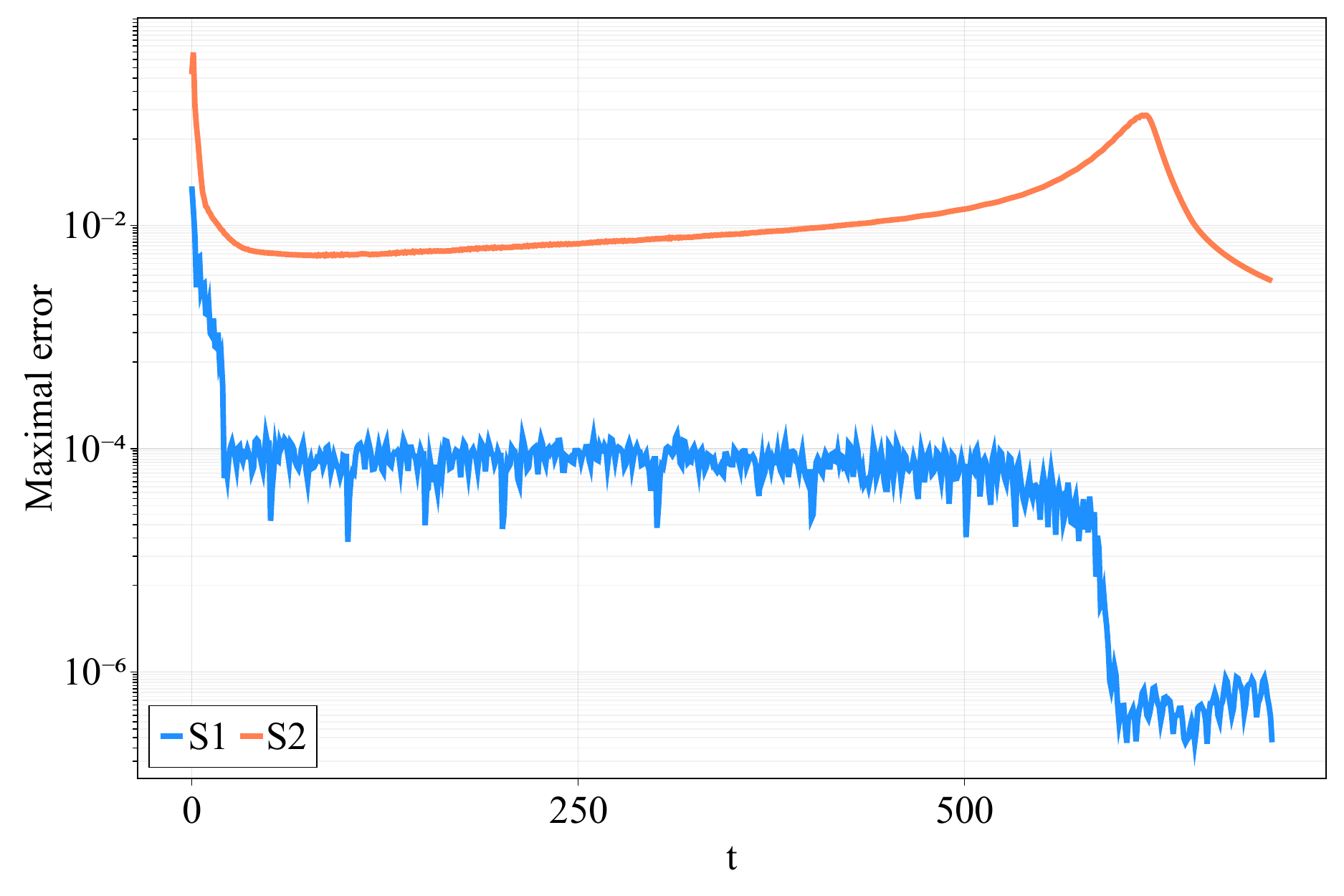}
\caption{The temporal evolution of the maximal numerical error at the black hole horizon in \textbf{S1} and \textbf{S2}.}
\label{fig1}
\end{figure}
\begin{figure}[h]
\centering
\includegraphics[width=9cm]{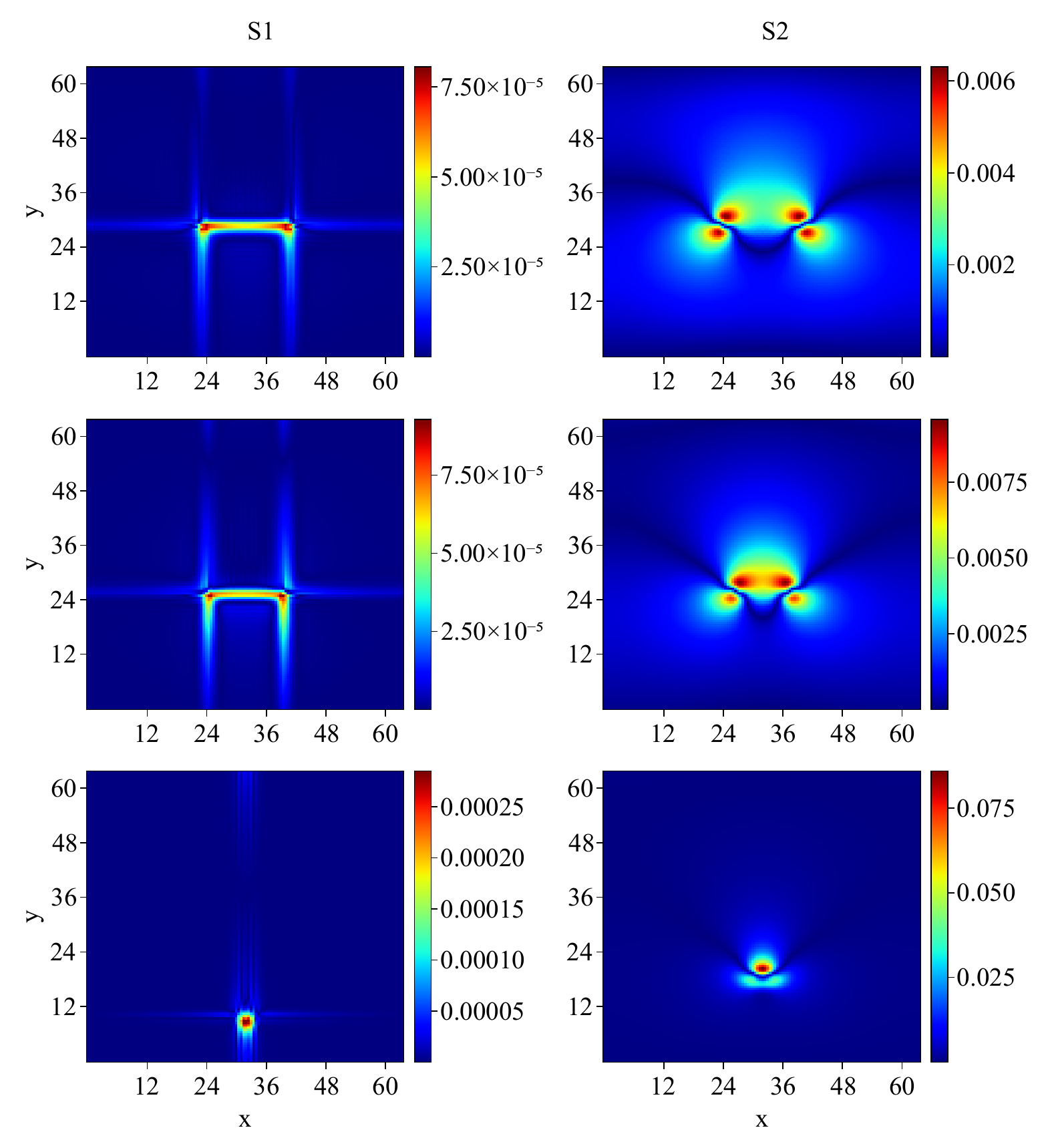}
\caption{The profile of $\mathcal{E}$ at the black hole horizon in \textbf{S1} on the left  and \textbf{S2} on the right, where the top panels are for the initial stage, the middle for the intermediate stage and the bottom for the annihilation stage. It turns out that the maximal error occurs in the neighbourhood of vortices.}
\label{fig3}
\end{figure}
\begin{figure}[h]
\centering
\includegraphics[width=9cm]{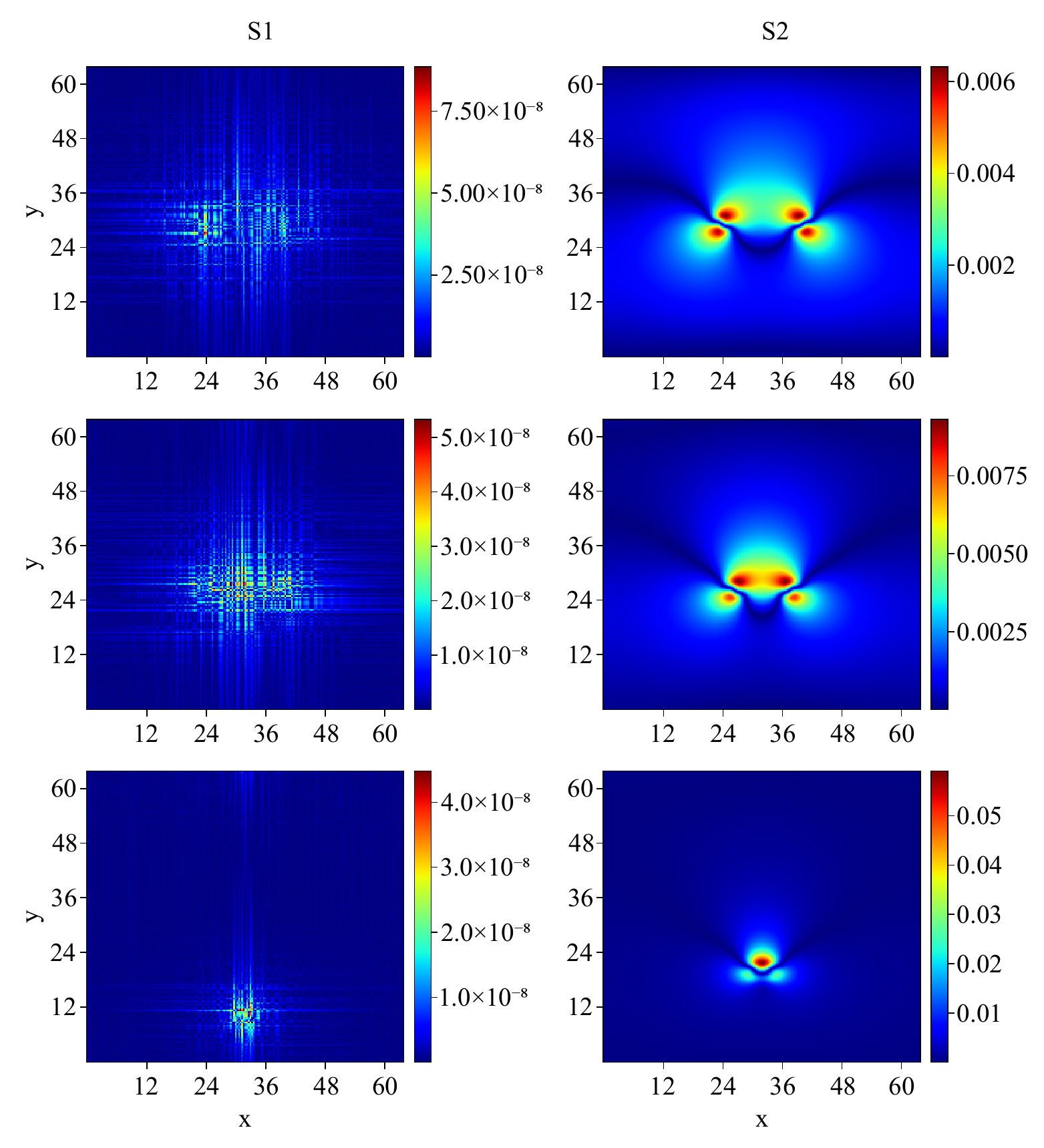}
\caption{The profile of $\mathcal{E}$ in \textbf{S1} on the left and \textbf{S2} on the right for the initial, intermediate and annihilation stages from the top to the bottom, where the number of grid points is taken as $192\times192\times64$.}\label{fignc}
\end{figure}

As a demonstration, we first plot the temporal evolution of the maximal error $\mathcal{E}_{max}$ at the horizon in Fig.\ref{fig1} for $\tilde{\mu}=4.5$, where the number of grid points in the $(x, y, z)$-direction is taken as $128\times 128\times 32$.  As one can see,
although $\mathcal{E}_{max}$ is appreciable in the beginning due to the aforementioned rough initial data, the black hole horizon offers a natural damping mechanism to make it die away quickly to order of $10^{-4}\sim10^{-5}$ in $\textbf{S1}$. In particular, when the vortex dipole annihilate, $\mathcal{E}_{max}$ is further decreased into order of $10^{-7}$, which indicates that the maximal error lies in the location of vortices. Such an observation is further confirmed in Fig.\ref{fig3}. While in $\textbf{S2}$, not only does $\mathcal{E}_{max}$ die down simply to order of $10^{-2}\sim10^{-3}$, but also stop to increase gradually during the later evolution. Moreover, it experiences a sharper increase right before the vortex dipole annihilation. This suggests that $\textbf{S2}$ is not as  applicable as \textbf{S1} for one to investigate the holographic superfluid dynamics, especially the vortex dynamics under consideration. In support of such a suspicion, we further examine the convergence of the error function by increasing the grid points. By comparing Fig.\ref{fig3} and Fig.\ref{fignc},  one can see that as the number of the grid points is increased, the numerical error in \textbf{S1} decreases dramatically while that in \textbf{S2} keeps almost unchanged.

So \textbf{S2} leads to the breakdown of
Eq.(\ref{eqAt}) in the whole bulk as well as the violation of the salient current conservation law on the boundary, which makes their results unreliable. In the main body of this Letter, we shall match the dissipative Gross-Pitaevskii equations with the holographic data obtained by our numerical evolution scheme.

\newpage
\subsection{The dispersion relation of DGPEs}
In this section, let us consider the dispersion relation of KGPE and LGPE. Firstly, considering the linear perturbations
\begin{equation}
\delta\psi=pe^{-i\omega t+i\textbf{k}\cdot\textbf{x}}+\bar{p}e^{i\omega^* t-i\textbf{k}\cdot\textbf{x}},\label{linper}
\end{equation}
on top of the static and homogeneous equilibrium configuration $\psi=1$,
where $p$ and $\bar{p}$ are independent of each other.  Then as usual, the dispersion relation can be obtained by substituting Eq.(\ref{linper}) into the linearized equation of motion as an eigenvalue problem. For KGPE, the resulting dispersion relation is
\begin{equation}
\omega(\textbf{k})=\pm\frac{\sqrt{\textbf{k}^4+4\textbf{k}^2\mu-4\gamma^2\mu^2}}{2\tau}-\frac{i\gamma(\textbf{k}^2+2\mu)}{2\tau}.\label{rootk}
\end{equation}
When $\gamma=0$, the long wavelength limit gives rise to the familiar dispersion relation for the sound mode. While in the presence of dissipation, we have
\begin{eqnarray}
    \omega_+(\textbf{k})&=&-i\frac{1}{2\tau\gamma}(1+\gamma^2)\textbf{k}^2, \nonumber\\
    \omega_-(\textbf{k})&=&-i[\frac{2\gamma\mu}{\tau}+\frac{1}{2\tau\gamma}(1-\gamma^2)\textbf{k}^2]
\end{eqnarray}
as $\mathbf{k}\rightarrow 0$. It is obvious that the sound mode behaves abnormally as the sound speed becomes zero.

Different from KGPE, the dispersion relation for LGPE produces the normal behavior of the sound mode as
\begin{equation}
\omega(\textbf{k})=\frac{\sqrt{\mu}}{\tau} \textbf{k}-\frac{i\lambda}{2\tau}\textbf{k}^2
\end{equation}
at small $\textbf{k}$.

\subsection{Good agreement between holography and LGPE for more complicated vortex motions}
\begin{figure}[h]
\centering
\subfigure[]{
  \includegraphics[width=8cm,height=5cm]{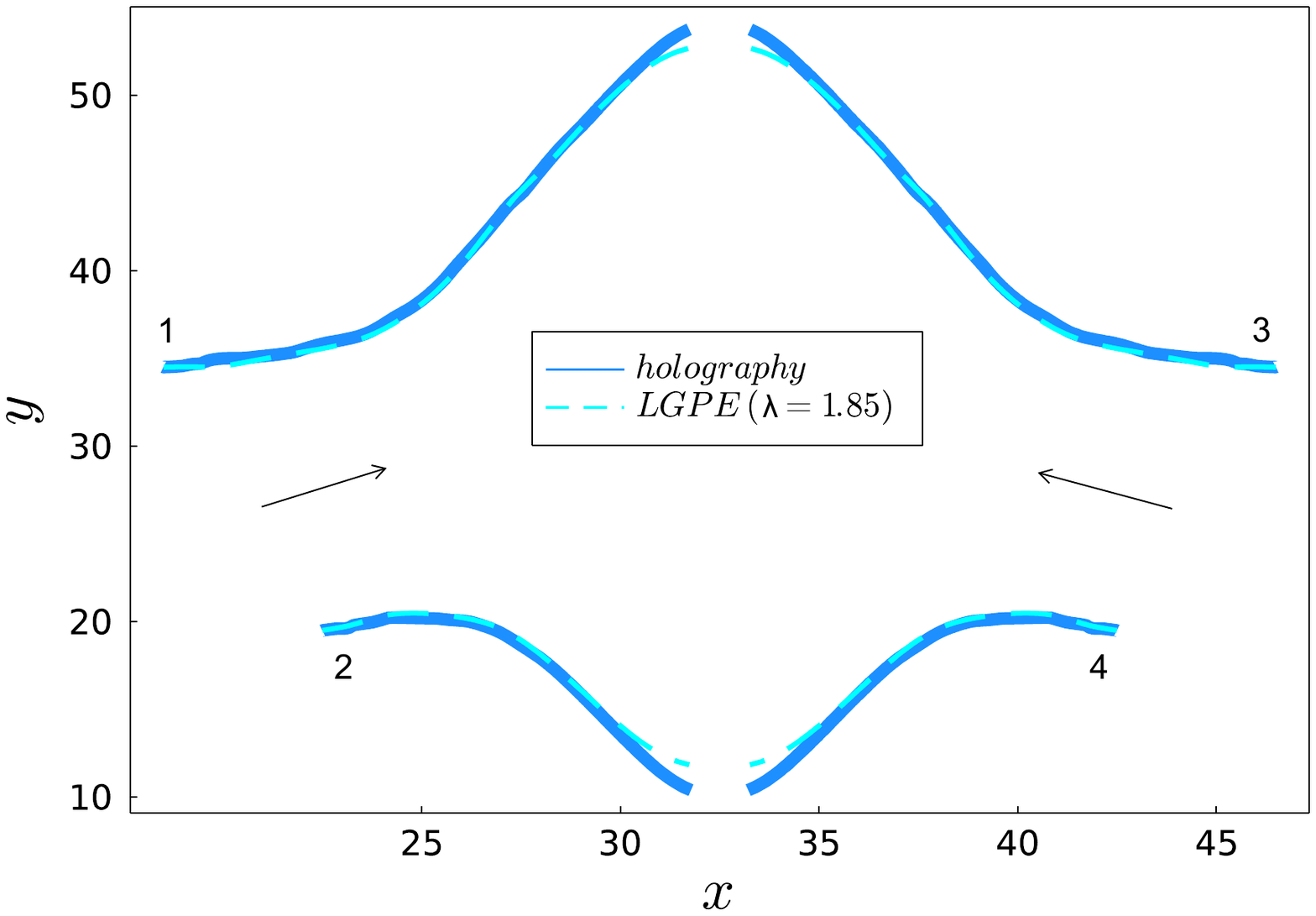}}
\subfigure[]{
  \includegraphics[width=8cm,height=5cm]{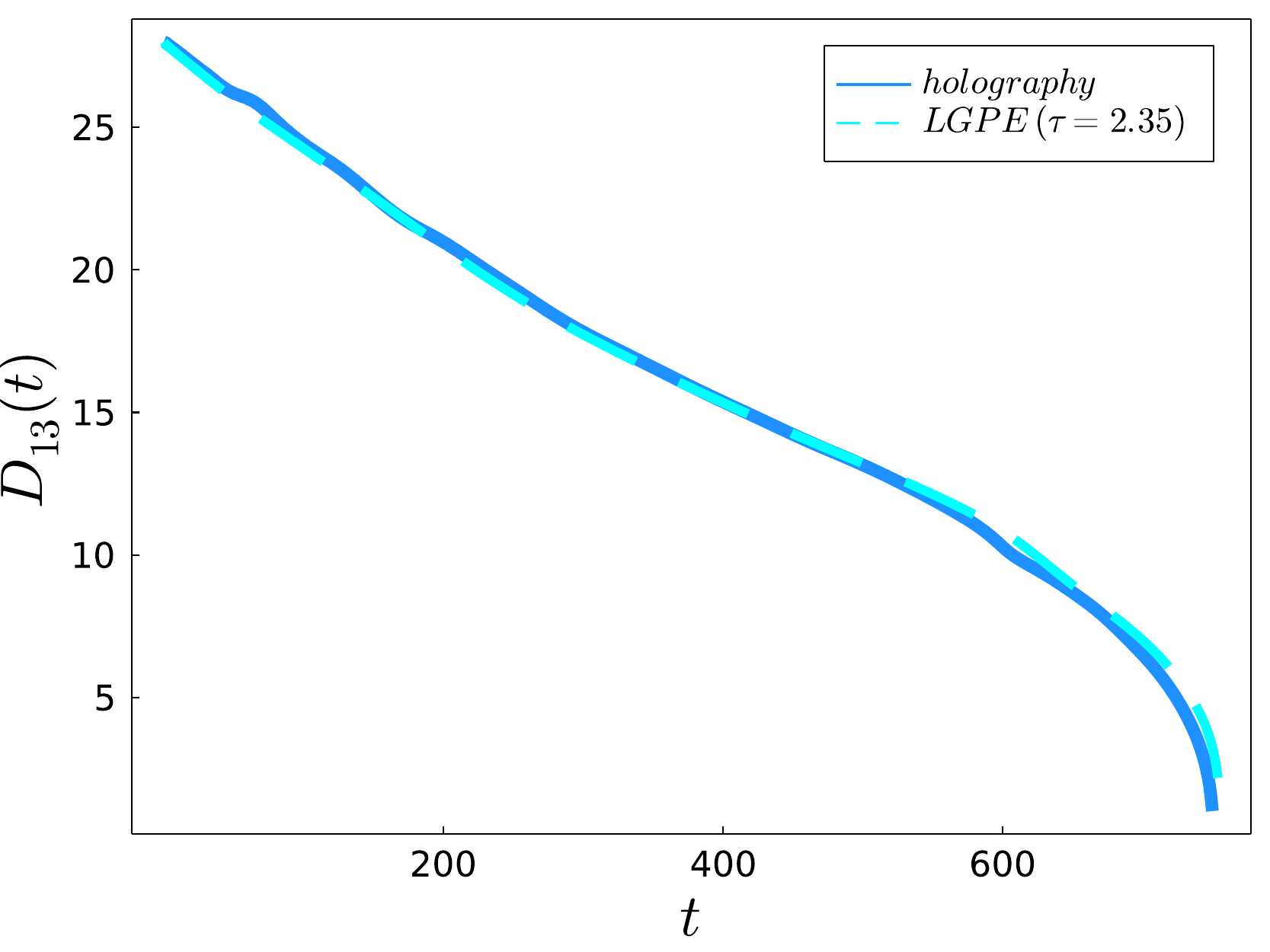}}
\caption{The good agreement between the holographic simulation at $\tilde{\mu}=4.5$ and the matched LGPE on the trajectories of vortices in (a)  and the temporal evolution of the relative distance between the vortex `1'  and the antivortex `3' from in (b) during the oblique collision of the two vortex dipoles.}\label{la}
\end{figure}

\begin{figure}[h]
\centering
\subfigure[]{
  \includegraphics[width=8cm,height=5cm]{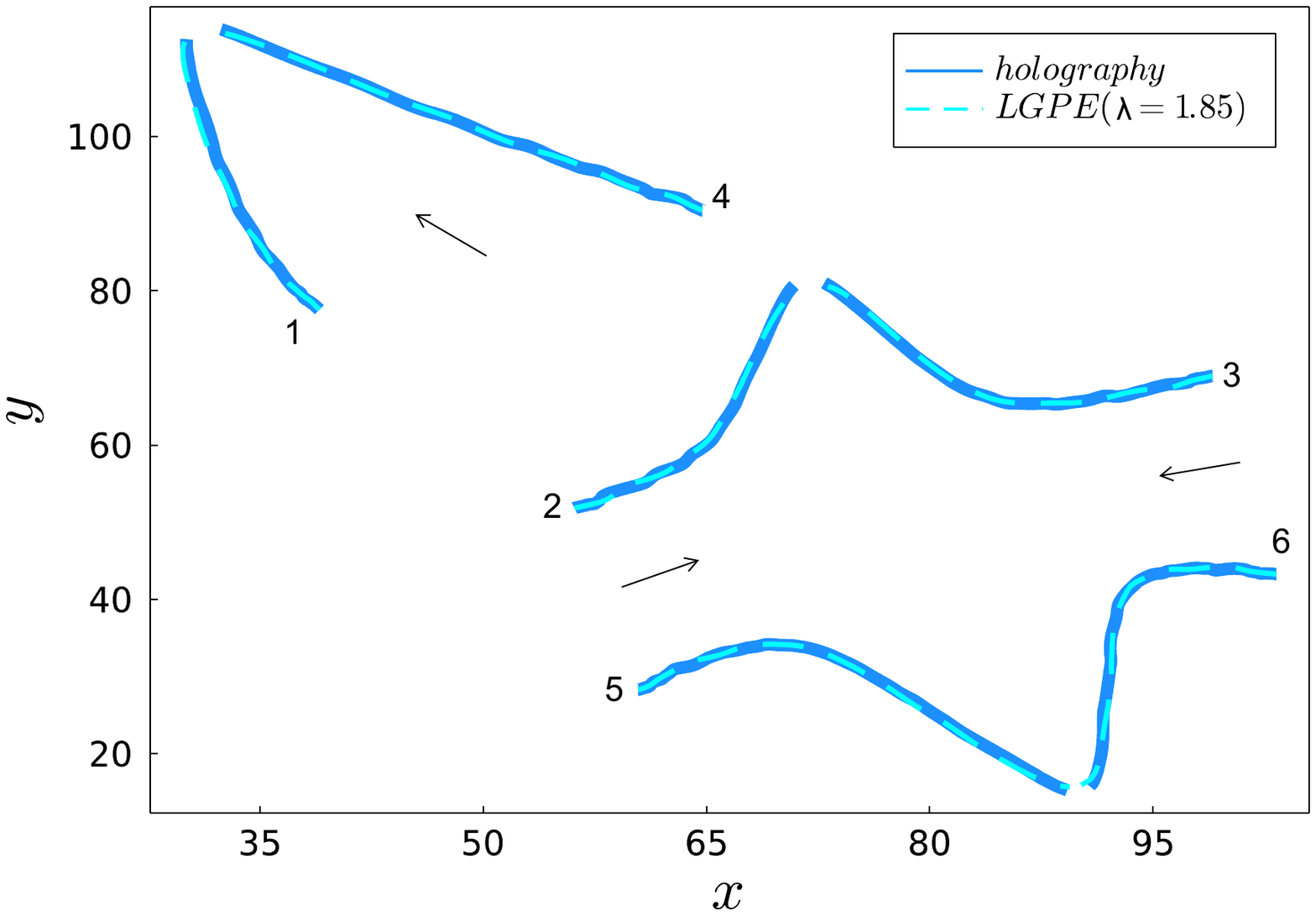}}
\subfigure[]{
  \includegraphics[width=8cm,height=5cm]{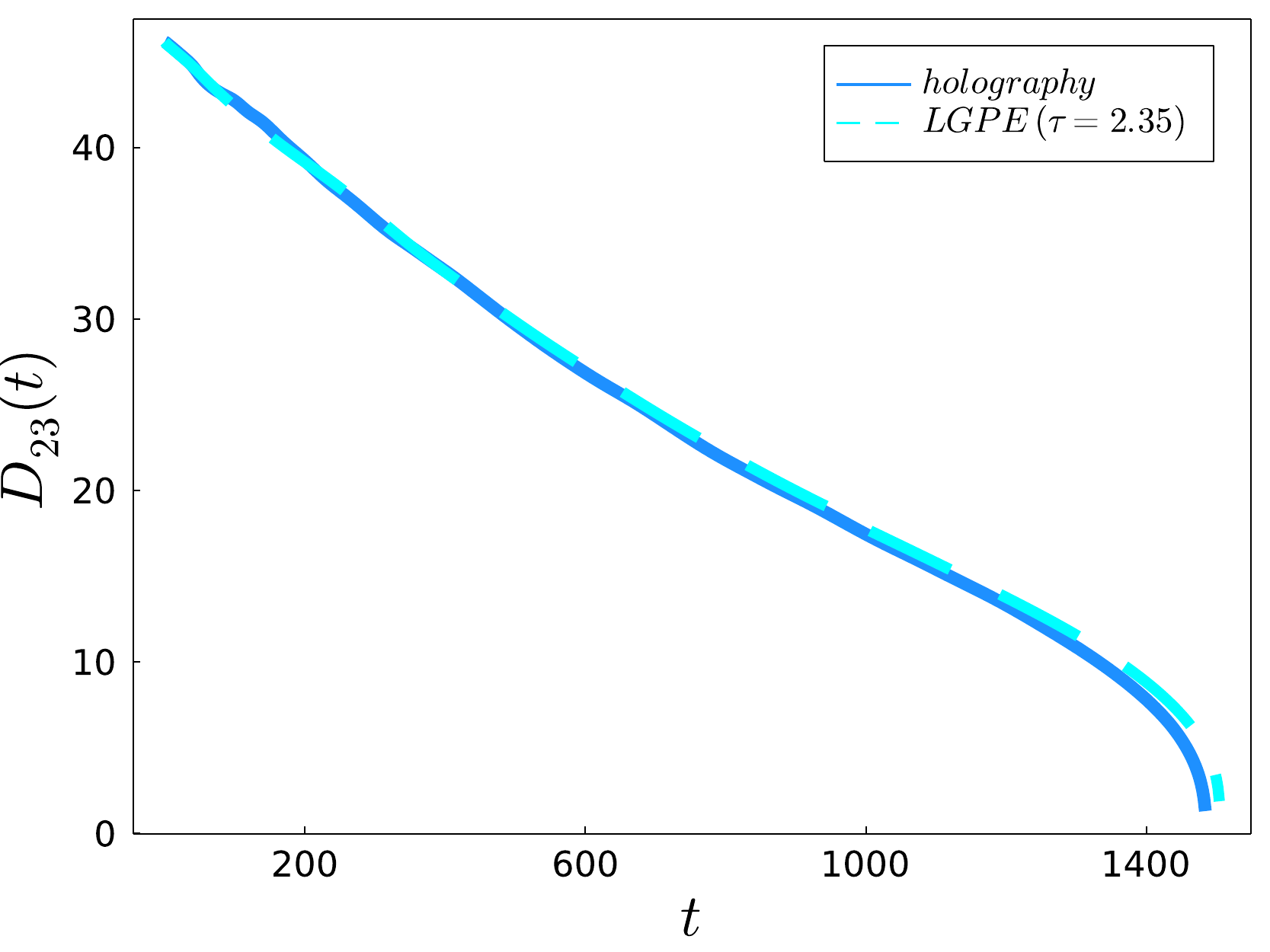}}
\caption{The good agreement between the holographic simulation at $\tilde{\mu}=4.5$ and the matched LGPE on the trajectories of vortices in (a)  and the temporal evolution of the relative distance between the vortex `2'   and the antivortex `3' in (b) for the random motion of six vortices, where the winding number of vortex `2', `4', and `6' is set to $1$ and the rest's winding number is set to $-1$.}\label{6v}
\end{figure}

Here we would like to provide further numerical evidences for LGPE as an effective description of holographic vortex dynamics by demonstrating good agreement between the holographic simulaiton and the matched LGPE for the oblique collision of two vortex dipoles and the random motions of six vortices in FIG.\ref{la} and FIG.\ref{6v}, respectively. As one can see, the good agreement still survives remarkably till the vortex pair annihilation in such more complicated scenarios involving vortex dynamics.


\end{document}